\documentclass[12pt]{article}
\usepackage{graphicx}
\usepackage{amsmath}
\usepackage{url}

\begin{document}

\title{Multicomponent bi-superHamiltonian KdV systems}

\author{Devrim Yaz{\i}c{\i} and Oya O\v{g}uz \\  Physics Department, Y{\i}ld{\i}z
Technical University, \\ Davutpasa Campus, Istanbul, Turkey   \and
\"{O}mer O\v{g}uz
\\ Physics Department, Bogazi\c{c}i University, \\ 80815 Bebek,
Istanbul, Turkey}

\maketitle

\begin{abstract}

It is shown that a new class of classical multicomponent super KdV
equations is  bi-superHamiltonian by extending the method for the
verification of graded Jacobi identity. The multicomponent
extension of super mKdV equations is obtained by using the super
Miura transformation.

\end{abstract}

\section {Introduction}
The theories of infinite dimensional super integrable systems have drawn a lot of
attention in the last two decades, for example, see \cite{kuper},\cite{kuper2}.
The research on the classical multicomponent integrable systems has also become
quite active more recently \cite{svin, omer,G-A, G-A1, karas, adler, ma}. In this
work we construct an extension of classical multicomponent Korteweg de Vries
(KdV)system to multicomponent superintegrable systems by employing a
bi-superHamiltonian formalism. Such systems are called super because they contain
both bosonic and fermionic fields. However there is no a supersymmetry
transformation between the fields as it is known from the case of 1-component
super KdV equations \cite{kuper1},\cite{G-O},\cite{mathiu1}. On the other hand
there also exist supersymmetric extensions of KdV equation, namely there exist
supersymmetry transformations but this bi-superHamiltonian systems have a nonlocal
nature \cite{mathiu},\cite{das},\cite{sokolov}.

We first introduce skew symmetric super Hamiltonian operators. It is shown that
they satisfy the graded Jacobi identity by using the method of prolongation
\cite{mathiu1},\cite{olver}. The set of multicomponent super integrable partial
differential equations are derived by introducing associated super Hamiltonians.
Furthermore introducing super Miura transformation, a multicomponent super
extension modified Korteweg de Vries (mKdV) system is obtained. The paper is
organized as follows. In section 2 we investigate the properties of super
Hamiltonian operators. It is shown that the second Hamiltonian operator satisfies
the graded Jacobi identity by means of constraint between the constant parameters
of the system while the first one does trivially. In section 3 the multicomponent
super evolution equations are derived from super Hamiltonians . In section 4 we
obtain the multicomponent super mKdV equations by using multicomponent super Miura
transformation. It is observed that the new systems are reduced to well known
systems of one component super evolution equations and in the vanishing fermionic
fields limit we get the multicomponent and  one component corresponding KdV
systems.

\section {The super Hamiltonian operators and Jacobi identity }
This section is devoted to study of properties of super
Hamiltonian operators. We first consider a set of fields
$\phi_{A}$ which contains both commuting and anticommuting fields
such as
\begin{equation}\label{1}
  \phi_{A}=\begin{pmatrix}
    _{u_\alpha} \\
    _{\xi_a} \
  \end{pmatrix}
\end{equation}
where $u_{\alpha}(x,t)$ is assumed a commuting (bosonic)field while $\xi_{a}(x,t)$
is an anticommuting (fermionic)field in 1+1 dimensions, $\alpha=1,2,...,m$ and
$a=1,2,...,n$. A $Z_{2}$ grading is  introduced such that $\tilde{p}(\phi)$ equals
to zero if $\phi_{A}$ is commuting  or one if it is anticommuting.

The evolution equation of a continuous dynamical Hamiltonian
system is given by
\begin{equation}\label{2}
  \partial_t\phi_A=\sum_{B}J_{AB}\frac{\delta\textit{\textbf{H}}}
  {\delta\phi_B}=\sum_{B}J_{AB}E_{B}(H),
\end{equation}
where $E_{B}$ is the Euler operator,
\begin{equation}\label{3}
  E_{B}=\sum_{k=0}^\infty(-\partial_{x})^k\frac{\partial}{\partial_x^k \phi_{B}}
\end{equation}
and $J$ is a certain differential operator and
$\textit{\textbf{H}}$ is a suitable functional. Functionals are
defined as modulo the integral of total derivative terms as
\begin{equation}\label{4}
  \textit{\textbf{F}}=\int{F[\phi_{B}]dx}
\end{equation}
where $F[\phi_{A}]$ is the element of the algebra of functions of
$x$, the fields $\phi_{A}(x)$ and their derivatives. The operator
$J$ defines a Poisson bracket as
\begin{equation}\label{5}
  \{\textit{\textbf{F}},\textit{\textbf{G}}\}=\sum_{AB}\int
  [J_{AB}E_B(G)]E_A(F)dx
\end{equation}
Here the ordering of the arbitrary functionals
$\textit{\textbf{F}}$ and $\textit{\textbf{G}}$ becomes important
for the graded systems. The fundamental Poisson bracket is
\begin{equation}\label{6}
  \{\phi_A(x),\phi_B(x')\}=J_{AB}\delta(x-x')
\end{equation}
that leads to the following expression for the evolution equation
\ref{2}
\begin{equation}\label{7}
  \partial_t\phi_A=\{\phi_{A},\textit{\textbf{H}}\}.
\end{equation}
$J$ is called a Hamiltonian operator if the Poisson bracket is
skew-symmetric as
\begin{equation}\label{8}
  \{\textit{\textbf{F}},\textit{\textbf{G}}\}=-(-1)^{\tilde{p}(\textit{\textbf{F}}).
  \tilde{p}(\textit{\textbf{G}})}\{\textit{\textbf{G}},\textit{\textbf{F}}\}
\end{equation}
where the grading ${\tilde{p}(\textit{\textbf{F}})}$ equals to zero (one) if an
arbitrary functional $\textit{\textbf{F}}$ is bosonic (fermionic) and satisfies
the Jacobi identity which can be given as vanishing the prolongation of an
evolutionary vector field, $\textit{v}_{J\Theta}$, associated to every Hamiltonian
$\textit{\textbf{H}}$ as follows \cite{mathiu1},\cite{olver},
\begin{equation}\label{9}
  pr\textit{v}_{J\Theta}(I)=0
\end{equation}
where I is the graded cosymplectic functional two-vector given as
\begin{equation}\label{10}
  I=\frac{1}{2}\sum_{A,B}\int J_{AB}\Theta_B\wedge\Theta_Adx
\end{equation}
here the set $\Theta_A=\{\theta_\alpha,\eta_a\}$ forms a basis of bosonic and
fermionic uni-vectors, dual to the one-forms $\{u_\alpha,\xi_a\}$ respectively.
Notice that
\begin{equation}\label{12}
  \Theta_A\wedge\Theta_B=-(-1)^{\tilde{A}.\tilde{B}}\Theta_B\wedge\Theta_A
\end{equation}
and $\Theta\wedge\Theta\neq0$ if $\Theta$ is fermionic.

We now introduce the super Hamiltonian  operators \\
i)First one:
\begin{equation}\label{13}
  J_{AB}^{(1)}=\begin{pmatrix}
    _{\delta_{\alpha\beta}\partial_x} & _{0} \\
    _{0} & _{\delta_{ab}} \
  \end{pmatrix}
\end{equation}
ii)Second one:
\begin{equation}\label{14}
  J_{AB}^{(2)}=\begin{pmatrix}
    _{j_{\alpha\beta}} & _{j_{\alpha b}} \\
    _{j_{a \beta}} & _{j_{ab}} \
  \end{pmatrix}
\end{equation}
where
\begin{equation}
    j_{\alpha\beta}=b_{\alpha\beta}\partial_x^3+2C_{\alpha\beta\gamma}u_{\gamma}\partial_x
    +C_{\alpha\beta\gamma}u_{{\gamma},x}
\end{equation}
\begin{equation}
    j_{\alpha b}=K_{\alpha bd}\xi_d\partial_x+L_{\alpha bd}\xi_{d,x}
\end{equation}
\begin{equation}
    j_{a\beta}=M_{a{\beta}d}\xi_d\partial_x+N_{a{\beta}d}\xi_{d,x}
\end{equation}
\begin{equation}
    j_{ab}=\Lambda_{ab}\partial_x^2+\Omega_{ab\gamma}u_{\gamma}
\end{equation}
where $u_{\alpha,x}=\partial_{x}u_{\alpha}$ and all coefficients apart from
$u(x,t)$ and $\xi(x,t)$ are constants. It is very easy to see that the operator
$J_{AB}^{(1)}$ is a Hamiltonian operator because it is skew-symmetric and Jacobi
identity is trivially satisfied, there are no variable coefficients in its
expression. On the other hand second operator $J_{AB}^{(2)}$, which is
skew-symmetric, contains x and t dependent coefficients. In order to show that it
is a Hamiltonian operator the graded Jacobi identity  should be satisfied. The eq.
(\ref{9}) for the second operator becomes
\begin{equation}\label{19}
  pr\textit{v}_{J\Theta}(I)=\frac{1}{2}\int pr\textit{v}_{J\Theta}(J_{AB}^{(2)})
  \Theta_B\wedge\Theta_Adx=0
\end{equation}
Here Einstein sum rule is employed and it will be used from now on. Eq.(\ref{19})
can be written as
\begin{eqnarray}\label{20}
  pr\textit{v}_{J\Theta}(I)=\frac{1}{2}\int[pr\textit{v}_{J\Theta}(j_{\alpha\beta})
  \theta_{\beta}\wedge\theta_{\alpha}+pr\textit{v}_{J\Theta}(j_{\alpha b})
  \eta_b\wedge\theta_{\alpha} \nonumber \\
  pr\textit{v}_{J\Theta}(j_{a\beta})
  \theta_{\beta}\wedge\eta_a+pr\textit{v}_{J\Theta}(j_{ab})
  \eta_b\wedge\eta_a]dx=0
\end{eqnarray}
On the other hand, in general,
\begin{equation}\label{21}
  pr\textit{v}_{J\Theta}(J_{AB}^{(2)})=\sum_{E,F,k}\partial_{x}^{k}
  (J_{EF}^{(2)}\Theta_{F})\frac{\partial}{\partial(\partial_{x}^{k}\phi_{E})}(J_{AB}^{(2)})
\end{equation}
Here $k=0,1$. Furthermore,
\begin{eqnarray}\label{22}
  pr\textit{v}_{J\Theta}(J_{AB}^{(2)})=\sum_{k}\{\partial_{x}^{k}
  (j_{\lambda\rho}\theta_{\rho})\frac{\partial}{\partial(\partial_{x}^{k}u_{\lambda})}
  (J_{AB}^{(2)})+ \partial_{x}^{k}
  (j_{\lambda e}\eta_{e})\frac{\partial}{\partial(\partial_{x}^{k}u_{\lambda})}
  (J_{AB}^{(2)})+ \nonumber  \\
  \partial_{x}^{k}
  (j_{d \rho}\theta_{\rho})\frac{\partial}{\partial(\partial_{x}^{k}\xi_{d})}(J_{AB}^{(2)})+
  \partial_{x}^{k}
  (j_{de}\eta_{e})\frac{\partial}{\partial(\partial_{x}^{k}\xi_{d})}(J_{AB}^{(2)})\}
\end{eqnarray}
By introducing
\begin{eqnarray}\label{224}
             C_{\alpha\beta\lambda} & = & C_{\beta\alpha\lambda}  \nonumber \\
                 \Omega_{ab\lambda} & = & \Omega_{ba\lambda} \\
\Omega_{ab{\lambda}}K_{{\lambda}cd} & = & \Omega_{ac{\lambda}}K_{{\lambda}bd}
\nonumber
\end{eqnarray}
and  by using eq.(\ref{22}) in the eq.(\ref{21}), we finally obtain
\begin{eqnarray}\label{23}
  pr\textit{v}_{J\Theta}(I) & = &\frac{1}{2}\int\{C_{\alpha\beta\lambda}b_{\lambda\rho}
  (\theta_{\rho,xxx}\wedge\theta_{\beta}\wedge\theta_{\alpha}
  -2\theta_{\rho,xx}\wedge\theta_{\beta,xx}\wedge\theta_{\alpha} \nonumber \\
                            &   &-2\theta_{\rho,xx}\wedge\theta_{\beta,x}
  \wedge\theta_{\alpha,x}) \nonumber \\
                            &   & +C_{\alpha\beta\lambda}C_{\lambda\rho\gamma}
  (u_{\gamma}\theta_{\rho,x}\wedge\theta_{\beta}\wedge\theta_{\alpha}
  +u_{\gamma,x}\theta_{\rho,}\wedge\theta_{\beta}\wedge\theta_{\alpha} \nonumber \\
                            &   & +4u_{\gamma}\theta_{\rho,x}\wedge\theta_{\beta,x}
  \wedge\theta_{\alpha}+2u_{\gamma,x}\theta_{\rho}\wedge\theta_{\beta,x}\wedge\theta_{\alpha}) \nonumber \\
                            &   & +M_{c{\beta}b}(L_{{\alpha}ac}+M_{a{\alpha}c}
  -N_{a{\alpha}c})\xi_{b}\eta_{a}\wedge\theta_{\beta,x}\wedge
  \theta_{\alpha,x} \nonumber  \\
                            &   & -N_{c{\beta}b}(K_{{\alpha}ac}-L_{{\alpha}ac}
  +N_{a{\alpha}c})\xi_{b,x}\eta_{a,x}\wedge\theta_{\beta}\wedge
  \theta_{\alpha} \nonumber  \\
                            &   & +[2C_{\alpha\beta\gamma}K_{{\gamma}ba}
  -M_{c{\beta}a}(K_{{\alpha}bc}-L_{{\alpha}bc}+N_{b{\alpha}c})]
  \xi_{a}\eta_{b,x}\wedge\theta_{\beta,x}\wedge\theta_{\alpha}  \nonumber \\
                            &   & +[2C_{\alpha\beta\gamma}L_{{\gamma}ba}
  -N_{c{\alpha}a}(M_{b{\beta}c}-N_{b{\beta}c}+L_{{\beta}bc})]
  \xi_{a,x}\eta_{b}\wedge\theta_{\beta,x}\wedge\theta_{\alpha}  \nonumber \\
                            &   &+[\Lambda_{ca}(K_{{\alpha}bc}+M_{b{\alpha}c})-
  6\Omega_{ba{\lambda}}b_{\lambda\alpha}]\eta_{a,xx}\wedge\eta_{b,x}\wedge
  \theta_{\alpha} \nonumber  \\
                            &   &+[\Lambda_{ca}(M_{b{\alpha}c}-N_{b{\alpha}c}
  +L_{{\alpha}bc})-2\Omega_{ba{\lambda}}b_{\lambda\alpha}]\eta_{a,xxx}\wedge\eta_{b}
  \wedge\theta_{\alpha} \nonumber  \\
                            &   &+[2\Omega_{ab{\lambda}}C_{\lambda\alpha\beta}+
  \frac{1}{2}\Omega_{ca{\beta}}(N_{b{\alpha}c}-4L_{{\alpha}bc}) \nonumber \\
                            &   &-\Omega_{cb{\beta}}M_{a{\alpha}c}]
  u_{\beta}\eta_{a}\wedge\eta_{b}\wedge\theta_{\alpha,x} \nonumber  \\
                            &   &+[\Omega_{ab{\lambda}}C_{\lambda\alpha\beta}-
  \frac{1}{2}\Omega_{ca{\beta}}(N_{b{\alpha}c}-2L_{{\alpha}bc})]
  u_{\beta,x}\eta_{a}\wedge\eta_{b}\wedge\theta_{\alpha} \nonumber  \\
                            &   &+\frac{1}{3}\Omega_{cb{\lambda}}[3L_{{\lambda}ad}
  -K_{{\lambda}ad}]\xi_{d,x}\eta_{a}\wedge\eta_{b}\wedge\eta_{c}\}dx=0
\end{eqnarray}

As it can easily be seen there is a trivial solution for eq.(\ref{23}) in which
all constant coefficients vanish. There exists a non-trivial solution as
\begin{eqnarray}\label{24}
   b_{\alpha\lambda}C_{\lambda\beta\gamma}     & = & b_{\beta\lambda}
   C_{\lambda\alpha\gamma}  \nonumber  \\
   C_{\alpha\beta\lambda}C_{\lambda\gamma\rho} & = & C_{\alpha\gamma\lambda}
   C_{\lambda\beta\rho}  \nonumber  \\
                               K_{{\lambda}ab} & = & M_{a{\lambda}b}  \nonumber  \\
                               K_{{\lambda}ab} & = & 3L_{{\lambda}ab}  \\
                              2M_{a{\lambda}b} & = & 3N_{a{\lambda}b}  \nonumber  \\
                    \Lambda_{bc}M_{a{\alpha}c} & = & 3b_{\alpha\beta}
   \Omega_{ab{\beta}}  \nonumber  \\
                   M_{c{\alpha}b}K_{{\beta}ac} & = & M_{c{\beta}b}
   K_{{\alpha}ac}  \nonumber  \\
           C_{\alpha\beta\gamma}K_{{\gamma}ab} & = & K_{{\alpha}ac}K_{{\beta}bc}  \nonumber
\end{eqnarray}
Thus $J_{AB}^{(2)}$ becomes a Hamilton operator with the set of equations
(\ref{24}).It describes the second Poisson structure.For KdV equation one can
easily show that sum of two Hamilton operators of bi-Hamiltonian structure is also
a Hamilton operator because one of the Hamilton operator ($J=\partial_{x}$)
trivially satisfies Jacobi Identity \cite{olver}. In our case
$J_{AB}^{(1)}+J_{AB}^{(2)}$ satisfies the graded Jacobi identity with the
condition
\begin{equation}\label{241}
  \Omega_{ab{\beta}}-M_{a{\beta}b}-\frac{1}{2}(K_{{\beta}ab}+L_{{\beta}ab}-
N_{a{\beta}b})=0
\end{equation}
and using our solution (24) this equation becomes
\begin{equation}\label{242}
  \Omega_{ab{\beta}}+N_{a{\beta}b}=2K_{{\beta}ab}.
\end{equation}
and furthermore,  we obtain
\begin{equation}\label{243}
  \Omega_{ab{\beta}}=4L_{{\beta}ab}
\end{equation}
Thus the Hamilton operators $J_{AB}^{(1)}$ and $J_{AB}^{(2)}$ constitute a super
Hamiltonian pair. We can now rewrite the second operator in terms of
$L_{{\lambda}ab}$ as
\begin{equation}
   J_{AB}^{(2)}=\begin{pmatrix}
  _{b_{\alpha\beta}\partial_{x}^{3}+C_{\alpha\beta\gamma}(u_{\gamma}\partial_{x}
  +\partial_{x}u_{\gamma})} & _{L_{{\alpha}bc}(2\xi_{c}\partial_{x}
  +\partial_{x}\xi_{c})} \\
  _{L_{a{\beta}c}(\xi_{c}\partial_{x}
  +2\partial_{x}\xi_{c})} & _{\Lambda_{ab}\partial_{x}^{2}+4L_{{\lambda}ab}u_{\lambda}}
\end{pmatrix}
\end{equation}
However the equations (24) and (25) provides information about algebra related to
the evolution equations. In the next section we shall derive the corresponding
evolution equations that are coupled multicomponent super KdV equations.

\section {The multicomponent super KdV systems }
Bi-Hamiltonian formalism suggests the existence of infinitely many
conserved quantities $\{H_{k}\}$ satisfying the recursion relation
\begin{equation}\label{25}
  \sum_{B}J_{AB}^{(2)}E_{B}(H_{k-1})=\sum_{B}J_{AB}^{(1)}E_{B}(H_{k})
\end{equation}
where $k=1,2,3,...$. These infinitely many conserved quantities provide an
extension of super KdV hierarchy to multicomponent super KdV hierarchy. We now
introduce the first two conserved quantities to obtain first member of evolution
equations as
\begin{equation}\label{26}
  H_{0}=\frac{1}{2}\int[-\delta_{\alpha\beta}u_{\alpha}u_{\beta}
  +\delta_{ab}\xi_{a}\xi_{b,x}]dx
\end{equation}
and
\begin{eqnarray}\label{27}
  H_{1} & = & \frac{1}{2}\int[-b_{\alpha\beta}u_{\alpha,x}u_{\beta,x}
  +C_{\alpha\beta\gamma}u_{\alpha}u_{\beta}u_{\gamma}  \nonumber  \\
        &   & -\Lambda_{ab}\xi_{a,x}\xi_{b,xx}+2K_{{\alpha}ab}u_{\alpha}
  \xi_{a}\xi_{b,x}]dx.
\end{eqnarray}
Then one can easily derive integrable super coupled integrable evolution
equations, which admit infinitely many conserved quantities due to the recursion
relations (\ref{25}), by using
\begin{equation}\label{28}
  \partial_t\phi_A=\sum_{B}J_{AB}^{(1)}E_{B}(H_{1})=\sum_{B}J_{AB}^{(2)}E_{B}(H_{0}).
\end{equation}
In this way we get the new class of integrable multicomponent super KdV equations
by using the equations (24) and (26) as follows
\begin{eqnarray}\label{29}
  u_{\alpha,t} & = & b_{\alpha\beta}u_{\beta,xxx}+3C_{\alpha\beta\gamma}
  u_{\beta,x}u_{\gamma}+K_{{\alpha}ab}\xi_{a}\xi_{b,xx}  \\
     \xi_{a,t} & = & \Lambda_{ab}\xi_{a}+K_{{\lambda}ab}(\xi_{b}u_{\lambda,x}
  +2u_{\lambda}\xi_{b,x})
\end{eqnarray}
In the bosonic limit when the fermionic variables vanish the system reduces to
multicomponent KdV  systems, known as degenerate Svinolupov system in which
$b_{\alpha\beta}$ is nondiagonalizable \cite{omer},\cite{G-A}. In this case,
1-component limit is the KdV equation. Furthermore if we choose the coefficients
$b_{11}=-1$, $\Lambda_{11}=-4$, $C_{111}=2$, $K_{111}=3$ satisfying the constraint
equations (24) and variables $u_{1}=u$ and $\xi_{1}=\xi$ eqs.(33-34) becomes
\begin{eqnarray}\label{32}
    u_{t} & = & -u_{xxx}+6uu_{x}+3\xi\xi_{xx} \\
  \xi_{t} & = & -4\xi_{xxx}+6u\xi_{x}+3u_{x}\xi .
\end{eqnarray}
These are super KdV equations given in references \cite{kuper1},\cite{G-O}. In
other words, our equations reduce to one of the known 1-component super KdV
equations which consists of one bosonic and one fermionic variables.

\section {The multicomponent super mKdV systems }
In this section we first introduce a super extension of Miura transformation using
the notation of previous sections. The multicomponent super Miura transformation
is
\begin{eqnarray}\label{33}
  u_{\alpha} & = & v_{\alpha,x}+\frac{1}{2}C_{\alpha\beta\gamma}v_{\beta}v_{\gamma}
  +K_{{\alpha}bc}\varepsilon_{b}\varepsilon_{c}  \\
     \xi_{a} & = & \varepsilon_{a,x}+\frac{1}{3}M_{a{\lambda}b}v_{\lambda}\varepsilon_{b}
\end{eqnarray}
where $v(x,t)$ and $\varepsilon(x,t)$ are new bosonic and fermionic variables,
respectively. The multicomponent super mKdV equations can be obtained from the
multicomponent super KdV equations by the multicomponent super Miura
transformations. This implies that any solution of the multicomponent super mKdV
equations gives a solution of the multicomponent super KdV equations through the
multicomponent super Miura transformations.When we substitute the transformation
(\ref{33}) into the eqs. (\ref{29}) we get the multicomponent super mKdV equations
as
\begin{eqnarray}\label{34}
       v_{\alpha,t} & = & -v_{\alpha,xxx}+\frac{3}{2}C_{\alpha\beta\gamma}
  C_{\beta\lambda\rho}v_{\lambda}v_{\rho}v_{\gamma,x}  \nonumber  \\
                    &   & +\frac{1}{8}K_{{\beta}mn}C_{\alpha\beta\gamma}
  (2v_{\gamma,x}\varepsilon_{m}\varepsilon_{n,x}
  +v_{\gamma}\varepsilon_{m}\varepsilon_{n,x}) \nonumber  \\
                    &   & +\frac{1}{4}K_{{\alpha}mn}\varepsilon_{n,x}\varepsilon_{m,xx} \\
  \varepsilon_{a,t} & = & -4\varepsilon_{a,xxx}-K_{{\beta}ab}
  (v_{\beta,xx}\varepsilon_{b}+2v_{\beta,x}\varepsilon_{b,x})  \nonumber  \\
                    &   & +K_{{\beta}ab}C_{\beta\lambda\rho}(v_{\lambda}v_{\rho}
  \varepsilon_{b,x}+v_{\lambda,x}v_{\rho}\varepsilon_{b})
\end{eqnarray}
by employing the constraints (\ref{24}) on the coefficients. As in the case of
multicomponent super KdV equations, the eqs.(39-40)reduces to
\begin{eqnarray}\label{35}
            v_{t} & = & \partial_{x}(2v^{3}-v_{xx}+\frac{3}{4}\varepsilon\varepsilon_{xx}
  +\frac{3}{2}v\varepsilon\varepsilon_{x}) \\
  \varepsilon_{t} & = & -4\varepsilon_{xxx}+(6vv_{x}-3v_{xx})\varepsilon
  +6(v^{2}-v_{x})\varepsilon_{x}
\end{eqnarray}
in the 1-component limit, by taking the coefficients as $C_{111}=2$, $K_{111}=3$
 and the variables as $v_{1}=v$ and $\varepsilon_{1}=\varepsilon$. This is the super
extension of the mKdV equation given by Kuperschmidt \cite{kuper1}.

\section {Conclusion}
In this work we have found the new class of integrable multicomponent super KdV
equations. It is shown that they are bi-superHamiltonian. The graded Jacobi
identity associated to the Poisson structure defined by super Hamiltonian
operators is satisfied by imposing constraints (\ref{24}) on the coefficients
introduced in the super Hamiltonian operators. These constraints could be
important to describe the structure associated to our evolution equations. It is
natural to expect that such relations would also imply the existence of
generalized symmetries. Furthermore by introducing a super Miura transformation a
super extension of multicomponent mKdV equations is obtained. This system also
possesses the structure described by the constraints (\ref{24}). We have shown
that our equations are reduced to well known 1-component super equations and
multicomponent and 1-component bosonic equations.



\begin{thebibliography}{10}
\bibitem{kuper} Kupershmidt B.A. "Elements of superintegrable systems"
D.Reidel,Dortrecht,(1987).
\bibitem{kuper2} Kupershmidt B.A.(Ed.) "Integrable and super integrable systems",
Singapore:World Scientific, (1990).
\bibitem{svin} Svinolupov S.I. Functional Anal.Appl.{\bf{27}},(1994),257.
\bibitem{omer} O\v{g}uz \"{O}."Hamiltonian structure of multicomponent KdV
equations" in "Symmetries in Science VIII" Ed. by B.Gruber, Plenum Press, New York
and London, (1995),405.
\bibitem{G-A} G\"{u}rses M. and Karasu A. Phys.Lett. 214{\bf{A}},(1996),21.
\bibitem{G-A1} G\"{u}rses M. and Karasu A. J.Math.Phys. {\bf{39}},(1998),2103.
\bibitem{karas} Karasu (Kalkanli) A. J.Math.Phys. {\bf{38}},(1997),3616.
\bibitem{adler} Adler V.E.,Svinolupov S.I.nad Yamilov R.I.Phys.Lett. 254{\bf{A}},(1999),24.
\bibitem{ma} Ma Wen-Xiu, J.Phys.A:Math.Gen. {\bf{31}}(1998), 7585.
\bibitem{kuper1} Kupershmidt B.A. Phys.Lett. 102{\bf{A}},(1984),213.
\bibitem{G-O} G\"{u}rses M. and O\v{g}uz \"{O}. Phys.Lett. 108{\bf{A}},(1985),437.
\bibitem{mathiu1} Mathieu P. Lett. in Math.Phys. {\bf{16}},(1988),199.
\bibitem{mathiu} Mathieu P. J.Math.Phys. {\bf{29}},(1988),2499.
\bibitem{das} Barcelos-Neto J. and Das A. J.Math.Phys. {\bf{33}},(1992),2743.
\bibitem{sokolov} Delduc F. and Gallot L. J.Math.Phys. {\bf{39}},(1998),4729.
\bibitem{olver} Olver P.J. "Applications of Lie Groups to Differential equations"
Springer-Verlag, New York, (1986).
\end{thebibliography}
\end{document}